\documentclass[usenatbib,useAMS]{mn2e}
\usepackage{graphicx}
\usepackage{epstopdf}
\usepackage{ctable}
\usepackage{url}
\usepackage{times}

\newcommand{\msun}{~\mathrm{M_{\odot}}}
\newcommand{\msunperyr}{~\mathrm{M_{\odot} {\rm ~yr}^{-1}}}

\newcommand{\mstar}{M_{\star}}
\newcommand{\mdust}{M_\mathrm{d}}

\newcommand{\acknowledgments}{\begin{small}\section*{Acknowledgments}\end{small}}

\title[Blended SMGs]{Spatially unassociated galaxies contribute significantly to the blended submillimetre galaxy population:~predictions
for follow-up observations of ALMA sources}

\author[C.~C. Hayward et al.]{
\parbox[t]{\textwidth}{
Christopher C. Hayward$^{1}$\thanks{E-mail: christopher.hayward@h-its.org},
Peter S. Behroozi$^{2,3}$, Rachel S. Somerville$^4$, Joel R. Primack$^5$, Jorge Moreno$^{6}$\thanks{CITA National Fellow}, 
and Risa H. Wechsler$^{2,3}$}
\vspace*{6pt} \\
$^1$Heidelberger Institut f\"ur Theoretische Studien, Schloss--Wolfsbrunnenweg 35, 69118 Heidelberg, Germany \\
$^2$Kavli Institute for Particle Astrophysics and Cosmology, Department of Physics, Stanford University, Stanford, CA 94305, USA \\
$^3$Department of Particle Physics and Astrophysics, SLAC National Accelerator Laboratory, Menlo Park, CA 94025, USA \\
$^4$Department of Physics \& Astronomy, Rutgers University, 136 Frelinghuysen Road, Piscataway, NJ 08854, USA \\
$^5$Department of Physics, University of California at Santa Cruz, Santa Cruz, CA 95064, USA \\
$^6$Department of Physics \& Astronomy, University of Victoria, 3800 Finnerty Road, Victoria, BC V8P 5C2, Canada}

\usepackage{hyperref}

\begin{document}

\date{Accepted for publication in MNRAS}

\pagerange{\pageref{firstpage}--\pageref{lastpage}} \pubyear{2013}

\maketitle

\label{firstpage}

\begin{abstract}
There is anecdotal evidence that spatially and physically unassociated galaxies blended into a single submillimetre (submm) source contribute
to the submm galaxy (SMG) population. This work is the first to theoretically predict the number counts of such sources.
We generate mock SMG catalogues using lightcones derived from the \textit{Bolshoi} cosmological simulation;
to assign submm flux densities to the mock galaxies, we use a fitting function previously derived from the results of dust radiative transfer performed
on hydrodynamical simulations of isolated disc and merging galaxies. We then calculate submm number counts for different beam sizes and without blending.
We predict that $\ga 50$ per cent of blended SMGs have at least one spatially unassociated component with $S_{850} > 1$ mJy.
For a 15-arcsec beam, blends of $>2$ galaxies in which at least one component is spatially unassociated dominate the blended sources
with total $S_{850} \ga 3$ mJy. The distribution of the redshift separations amongst the components is strongly bimodal.
The typical redshift separation of spatially unassociated blended sources is $\sim 1$.
Our predictions for the contributions of spatially unassociated components and the distribution of redshift separations are
not testable with currently available data, but they will be easily tested once sufficiently accurate redshifts for the individual subcomponents (resolved by, e.g., ALMA)
of a sufficient number of single-dish-detected blended SMGs are available.
\end{abstract}


\section{Introduction} \label{S:intro}

Submillimetre (submm) galaxies\footnote{Throughout this work, we use the acronym SMG in the conventional manner -- to denote single-dish-detected submm
sources -- although a given SMG may be composed of multiple distinct galaxies. This issue is discussed in Section \ref{S:importance_mock_obs}.}
(SMGs; see \citealt{Blain:2002} for a review) are some of the most luminous, rapidly star-forming galaxies (with star formation rates of
$\sim 10^2 - 10^3 \msunperyr$; e.g., \citealt{Magnelli:2010,Magnelli:2012,Chapman:2010,Michalowski:2010masses,Michalowski:2010production}) in the Universe.
Although simulations have suggested that merger-induced starbursts can produce the submm fluxes characteristic of SMGs \citep{Chakrabarti:2008SMG,Narayanan:2009,Narayanan:2010dog,
Narayanan:2010smg,Hayward:2011smg_selection}, explaining the observed number density
of such prodigious star-formers has posed a challenge for semi-analytic models and hydrodynamical simulations
(e.g., \citealt{Granato:2000,Baugh:2005,Fontanot:2007,Dave:2010}; \citealt{Hayward:2011num_cts_proc,Hayward:2013number_counts}; \citealt*{Narayanan:2012gas_frac}; \citealt*{Narayanan:2012IMF};
\citealt{Niemi:2012}; \citealt*{Shimizu:2012}; \citealt{Somerville:2012}), which has caused some to consider them a challenge for $\Lambda$CDM \citep{Primack:2012}.

Recently, \citet{Hayward:2011smg_selection,Hayward:2012smg_bimodality,Hayward:2013number_counts} suggested that treatment of the effects of blending caused by the large
beam sizes (FWHM $\sim 15$ arcsec, or $\sim 130$ kpc at $z \sim 2-3$) of the single-dish telescopes used to identify SMGs can help alleviate the discrepancy between the SMG number
counts predicted by models and those observed. Specifically, during the pre-coalescence phase of major mergers, both progenitor discs can be blended
into a single submm source; \citet[][hereafter, H13]{Hayward:2013number_counts} suggest that such `galaxy-pair' SMGs account for a significant fraction ($\ga 30$ per cent) of
the SMG population. Interferometric observations \citep[e.g.,][]{Tacconi:2006,Tacconi:2008,Bothwell:2010,Riechers:2011a,Riechers:2011b,
Engel:2010,Wang:2011,Barger:2012,Smolcic:2012,Karim:2013} have provided many
examples of single-dish SMGs that are resolved into multiple components when observed at an order-of-magnitude better resolution, and in some cases,
the redshifts and kinematics suggest that the individual sources are widely separated (projected separation $\ga 10$ kpc) discs in the process of merging.
Recent interferometric continuum imaging surveys suggest that of order half of single-dish-detected SMGs
are blends of two or more distinct components \citep{Smolcic:2012,Karim:2013}. Interestingly, in those surveys, the brightest sources (those with 870-\micron~flux
density $S_{870} \ga 12$ mJy) are almost all blended sources. However, there are at least fifteen examples of interferometrically observed sources that are brighter than the suggested
flux density cutoffs (e.g., \citealt{Dannerbauer:2002,Younger:2007high-z_SMGs,Younger:2008LH,Younger:2008,Younger:2009SMG_interf,Barger:2012,Hodge:2012,
Smolcic:2012CARMA,Wagg:2012};
see \citealt{Hayward:2013limits} for further details). Thus, the fraction of single-dish submm sources that are blended and the brightest submm flux density of a
single galaxy are still uncertain.

A submm flux-density cutoff suggests an upper limit on the star formation rate (SFR) of SMGs
\citep{Karim:2013,Hayward:2013limits}, which may be a consequence of feedback from star formation and/or active galactic nuclei but may also be simply a
consequence of limited gas supply. It also suggests that to reproduce the brightest SMGs, models must account for the effects of blending. In this work, one question
that we address is whether we can reproduce the number counts of the brightest SMGs via blended sources alone.

Early-stage mergers are only one type of blended submm source; it is also possible that spatially (and thus physically) unassociated
projected multiples contribute to the SMG population. There is already anecdotal evidence for this subpopulation: \citet{Wang:2011} present two
examples of SMGs that are resolved into multiple components by the Submillimeter Array (SMA) and for which the
resolved components are located at significantly different redshifts (in one case, the difference between the two components is $\Delta z = 1.015$, and
in the other, the three components have redshifts of 2.914, 3.157, and 3.46). Thus, the individual components are not widely separated galaxies that will merge
but rather chance projections of completely unrelated galaxies. Some of the sources observed by \citet{Smolcic:2012} may also be examples of this subpopulation.

The relative contribution of such projected multiples to the SMG population is currently unconstrained. However, this situation is likely to change soon,
once redshifts for a significant number of the sources from the Atacama Large Millimeter/submillimeter Array (ALMA) follow-up observations
\citep{Karim:2013,Hodge:2013} of the Large APEX Bolometer Camera (LABOCA; \citealt{Siringo:2009}) Extended \textit{Chandra} Deep
Field South Submillimetre Survey (LESS; \citealt{Weiss:2009}) sources are available. Thus, a prediction for
the relative contributions of spatially associated and unassociated components is particularly timely.

In this work, we make such a prediction. To do so, we use halo catalogues derived from a cosmological simulation. Stellar masses and SFRs are assigned
using observationally constrained stellar mass-- and SFR--halo mass relations.
For each galaxy, we assign an 850-\micron~flux density $S_{850}$ using a fitting function, which was previously derived from the results of dust radiative transfer performed
on hydrodynamical simulations of isolated disc and merging galaxies, that gives $S_{850}$ as a function of SFR and dust mass.
Then, we search the mock catalogues for blended submm sources and calculate the total submm flux density for each blended source. Using a cut based on the
redshift separations of the individual components of the blended sources, we separate
the blended sources into `spatially associated' and `spatially unassociated' subpopulations and calculate the relative contributions to the submm number counts as
a function of single-dish submm flux density.

The remainder of this work is organised as follows: in Section 2, we discuss the details of our method. In Section 3, we present our predictions for the cumulative submm number
counts for different beam sizes, the relative contributions of spatially associated and unassociated sources, and the distribution of redshift separations amongst the individual
components of blended sources. In Section 4, we discuss some implications of our results. Section 5 presents our conclusions and plans for future work.

\section{Methods} \label{S:methods}

\begin{figure*}
\centering
\includegraphics[width=\columnwidth]{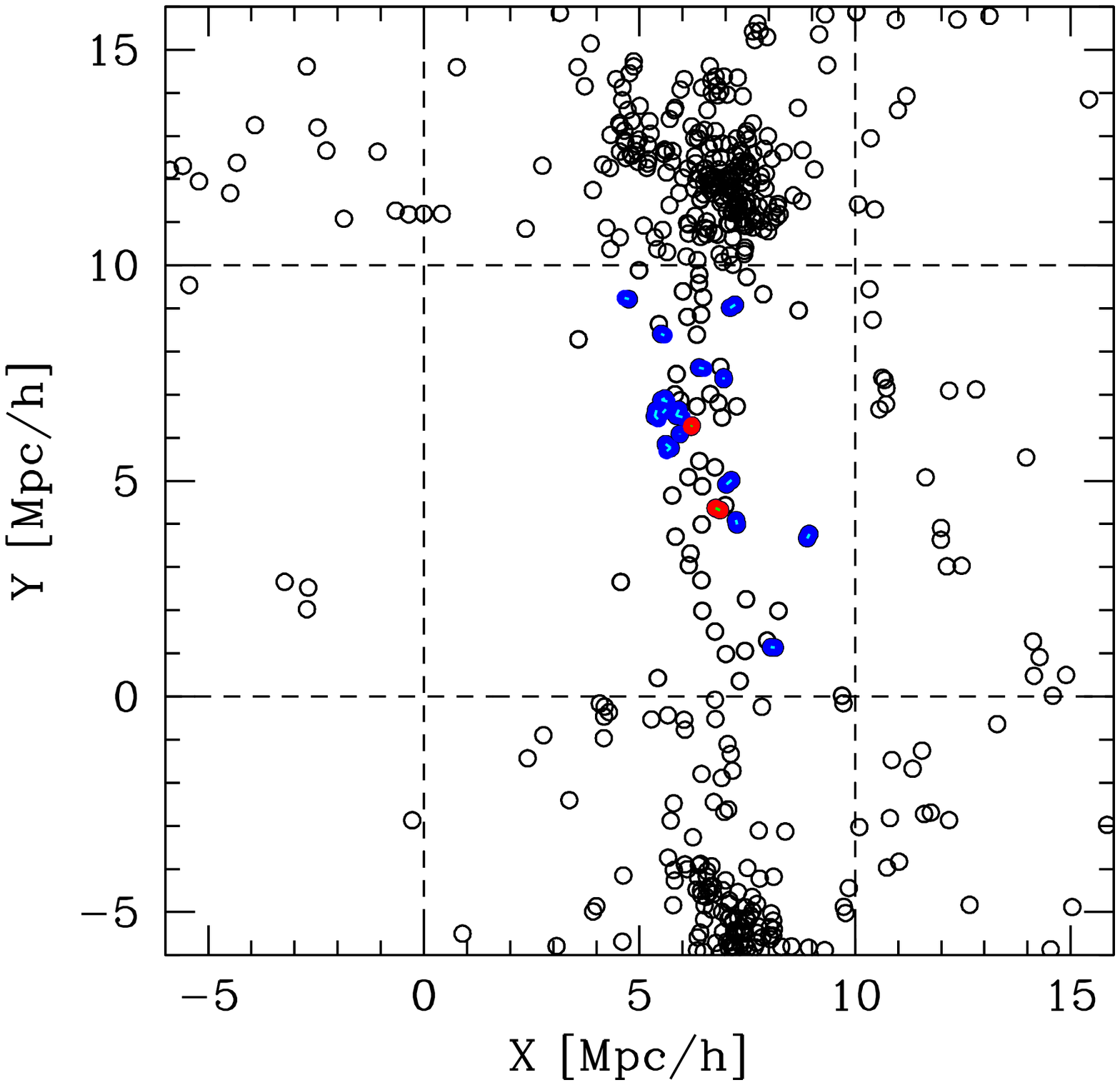}
\includegraphics[width=\columnwidth]{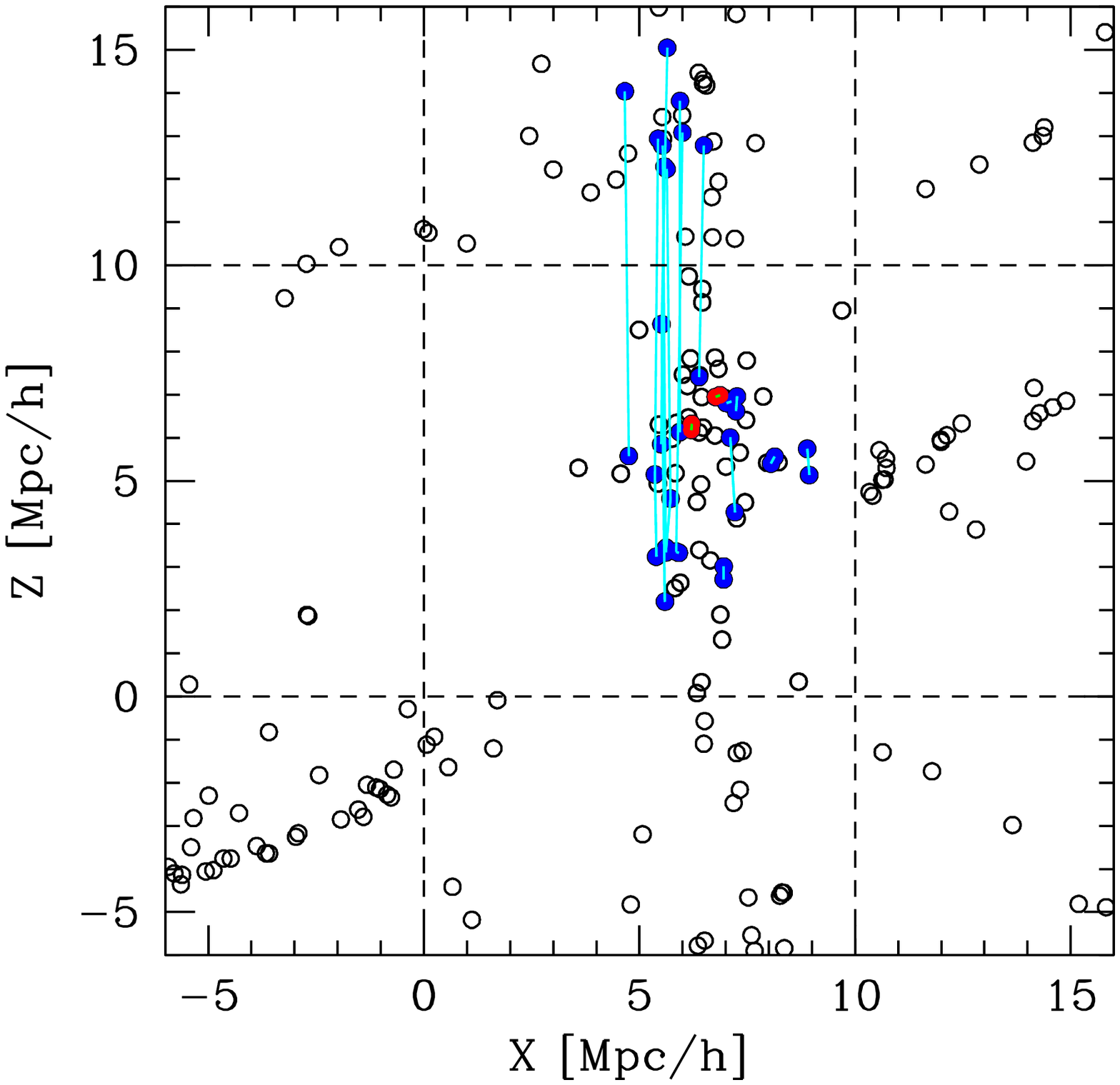}
\caption{Demonstration of the difference between spatially associated and spatially unassociated galaxy pairs. Open circles denote the locations of mock galaxies, and
dashed lines demarcate cubic regions with side length 10 $h^{-1}$ Mpc (comoving). The focus is on galaxies in the central cube and their neighbours (which may be located in the
neighbouring cubic regions). \textit{Left:} $X$-$Y$ plane. Red solid symbols denote galaxies that are associated in three dimensions, whereas blue solid symbols denote
galaxies that appear to be associated in projection but are actually unassociated in three dimensions. An observer without access to accurate redshifts would be unable to distinguish between
these two cases. \textit{Right}: $X$-$Z$ plane. This rotated perspective verifies that the galaxy pairs denoted with red symbols are associated, whereas those denoted with
blue symbols are certainly not. The lengths of the cyan lines connecting the unassociated galaxy pairs are substantially larger than their apparent separations in the projected $X-Y$ plane.}
\label{fig:false_pairs}
\end{figure*}

\subsection{Creating mock galaxy catalogues}

The method used in this work differs significantly from that of H13 because the method used in H13 cannot treat spatially unassociated projected multiples. Here,
we use mock lightcones derived from a cosmological simulation to provide information about the large-scale clustering of dark matter haloes.
We start with halo catalogues from the \textit{Bolshoi} simulation (\citealt*{Klypin:2011}; \citealt{Behroozi:2013rockstar,Behroozi:2013}),
which is a collisionless dark matter simulation performed using the Adaptive Refinement Tree (ART; \citealt*{Kravtsov:1997}) code.
The \textit{Bolshoi} simulation volume has side length 250 $h^{-1}$ Mpc and contains $2048^3$ particles, which yields a mass resolution of
$1.35 \times 10^8 h^{-1} \msun$. The force resolution is 1 $h^{-1}$ kpc (physical). The cosmological parameters used for \textit{Bolshoi} are
$\Omega_m = 0.27$, $\Omega_{\Lambda} = 0.73$, $h = 0.7$, $\sigma_8 = 0.82$, and $n_s = 0.95$. These parameters are consistent
with the WMAP7+BAO+H$_0$ results \citep{Komatsu:2011}. The values for $\sigma_8$ and $n_s$ are also consistent with the recent
\textit{Planck} results \citep{Planck:2013overview,Planck:2013parameters}. The value of $\Omega_m$ ($\Omega_{\Lambda}$ and $h$) used is
$2.6 \sigma$ greater ($2.2 \sigma$ less) than that from \textit{Planck}. Nevertheless, the uncertainty in the cosmological parameters
should affect our results less than the other uncertainties discussed below.

Starting at eight random locations within the simulation
volume, we select haloes along a randomly oriented sightline with an 84' x 84' (1.96 deg$^2$) field of view from $z=0.5$ to $z=8$; this redshift range contains the bulk of the SMG population,
both in the real Universe \citep[e.g.,][]{Smolcic:2012,Yun:2012,Michalowski:2012,Weiss:2013} and in our model (for reasons discussed below). Cosmological redshifts are calculated including
the effects of halo peculiar velocities. Stellar masses are assigned based on the stellar mass---halo mass relation of \citet*{Behroozi:2013SFH}, which accounts for redshift-dependent scatter in
stellar mass at fixed halo mass and the effects of systematic observational biases in stellar mass recovery. SFRs are assigned based on the SFR---halo mass relation of
\citet{Behroozi:2013SFH}; this includes redshift-dependent scatter in SFRs at fixed halo mass and the increasingly tight correlation between stellar mass and SFR at fixed halo mass at high redshifts.
To approximate environmental effects near massive clusters, satellite SFRs are reduced by a factor equal to their current subhalo mass divided by the peak mass in their subhalo's
mass accretion history. This is only a toy model for satellite quenching, of course. However, our results should be insensitive to the inclusion of satellite quenching
because satellites are typically too submm-faint to contribute to the (detectable) SMG population.

Our method ensures that the mock sample has stellar mass functions, average specific SFRs, cosmic star formation history, and
autocorrelation as a function of stellar mass that are consistent with observations. Because the model reproduces the mean SFR for a given stellar mass (including some scatter)
and redshift rather than $\textrm{SFR}(\mstar,z)$ for individual haloes, the elevation in SFR associated with starbursts is not included, so we simply do not treat that subpopulation of SMGs in
this work. This issue is discussed in detail in Section \ref{S:limitations}.

\subsection{Assigning submm flux densities}

To assign submm fluxes to the mock galaxies, we use the following fitting function, which was derived using the results of \textsc{Sunrise} \citep{Jonsson:2006sunrise,Jonsson:2010sunrise}
dust radiative transfer calculations performed on \textsc{Gadget-3} \citep{Springel:2005gadget} smoothed-particle hydrodynamics simulations of isolated disc and merging galaxies (eq. 15 of H13):
\begin{equation}
S_{850} = 0.81 {\rm ~mJy} \left(\frac{\mathrm{SFR}}{100 ~\msunperyr}\right)^{0.43} \left(\frac{\mdust}{10^8 \msun}\right)^{0.54},
\end{equation}
where $S_{850}$ is the observed-frame 850-\micron~flux density and $\mdust$ is the dust mass (we describe the method for assigning $\mdust$ below).
This function can recover the submm flux of simulated galaxies in the redshift range $z \sim 1-6$ to within a scatter of 0.13 dex because the negative $K$-correction makes the
observed-frame $S_{850}$ of a fixed galaxy SED almost independent of $z$ in the range $z \sim 1-10$ \citep[e.g.,][]{Blain:2002}. It under-predicts the flux at lower redshifts, significantly so for
$z \la 0.5$. However, because the normalization of the SFR--$\mstar$ relation decreases strongly from $z \sim 1$ to $z \sim 0$ \citep[e.g.,][]{Noeske:2007a,Daddi:2007}
and the volume in the range $z = 0-1$ is a small fraction of the total $z = 0-6$ volume, both the observed \citep[e.g.,][]{Smolcic:2012,Yun:2012,Weiss:2013} and predicted
(\citealt{Hayward:2012thesis}; H13) fraction of SMGs with $z \la 1$ is small; thus, this under-prediction should not significantly affect our results. Because the \citet{Eddington:1913}
bias (see also \citealt{Hogg:1998}) can affect the shape of the number counts, we incorporate the scatter of 0.13 dex when assigning flux densities to the mock galaxies.

The fitting function requires the SFR and $\mdust$ of each mock galaxy. The SFR is already determined from the \citet{Behroozi:2013SFH} method.\footnote{We have checked 
the effect of calculating the SFR following H13 instead. For the majority of the sources, the SFRs and thus submm flux densities are very similar. Consequently, the results
presented here are qualitatively insensitive to this choice, although there can be some quantitative discrepancies at the bright end because of the small number of sources
that contribute there.}
To assign dust mass values, we follow the method of H13, in which the gas fraction $f_g$ and metallicity $Z$ are parameterized as functions of stellar mass and redshift
and it is assumed that 40 per cent of the metals are contained in dust grains. The functional form for $f_g(\mstar,z)$ is physically motivated, and the parameters are constrained
to match $z \la 3$ observations; see
\citet{Hopkins:2010IR_LF} for details. The form for $Z(\mstar,z)$ is based on the observed stellar mass--metallicity relation at different redshifts; see H13 for details. We note
that because $S_{850}$ scales significantly sublinearly with both SFR and $\mdust$, the predictions are relatively insensitive to variations in these quantities; e.g., a factor of
2 change in either quantity changes $S_{850}$ by less than 50 per cent.

\subsection{Identifying blended sources}

Once submm flux densities are assigned to all mock galaxies, we perform a brute-force search for blended multiples. First, we select only those sources with
$S_{850} > 1$ mJy because fainter sources are below the flux density limits of current single-dish surveys and interferometric follow-up observations
(e.g., \citealt{Karim:2013}).\footnote{Sources fainter than the detection limit cause the zero-level of the submm map to be biased. However,
this bias can be corrected when deriving number counts from submm maps (see, e.g., section 3.2.3 of \citealt{Weiss:2009}), so we consider it reasonable to
ignore the contribution of such sources to the predicted number counts. Using a different flux density limit in our model naturally changes the number of components of individual sources because
for a lower (higher) flux density limit, the interferometric follow-up observations would detect more (fewer) components. We have checked the effects of using
alternate flux density limits of $S_{850} = 0.5$ and 2 mJy. The quantitative predictions for the absolute number counts change, but the relative contributions
of the subpopulations and the distributions of the redshift separations of components are relatively robust. Furthermore, the conclusions are qualitatively unchanged, and this uncertainty
is subdominant to that associated with the treatment of blending.} This flux cut yields a minimum flux density of $n$ mJy for $n$-component blended sources.
Then, for each source $i$ in this subsample, we compute the angular distance to all sources $j$ in the subsample for which $j > i$ (to avoid double-counting).
If one or more sources is within a `beam size'\footnote{Throughout this work, we refer to the `beam' and `beam size' used in our model, but it should be understood
that we are not claiming to precisely mimic the effects of blending (cf. fig. 1 of \citealt{Hayward:2012smg_bimodality}). Rather, our treatment is simply an approximation.
This limitation is discussed in Section \ref{S:limitations}.}
of source $i$, which we vary, we consider the sources to be blended and sum their $S_{850}$ values to calculate the total $S_{850}$ for the blended source.

For each blended source, we compute a measure of the redshift separation of the components by summing the redshift separations between
the first component and all other components in quadrature:
\begin{equation} \label{eq:Delta_z}
\Delta z \equiv \left(\sum^N_{i > 1} (z_i - z_1)^2\right)^{1/2},
\end{equation}
where $z_i$ denotes the redshift of the $i$th component and $N$ is the total number of components of the blended source. This is a somewhat
arbitrary (yet natural) measure: for two components, $\Delta z$ is simply the difference in the redshifts of the two components.
For sources with $>2$ components, $\Delta z \gg 0$ indicates that at least one of the components is at significantly different $z$ than the others.
Below, we define `spatially associated' and `spatially unassociated' blended sources by $\Delta z < 0.02$ and $\Delta z \ge 0.02$, respectively.
The justification for and insensitivity of our results to the value of this cut are presented in Section \ref{S:z_sep}.

Fig. \ref{fig:false_pairs} presents a simple illustration of the difference between spatially associated and unassociated projected pairs. The panels
show the positions of a sample of mock galaxies projected onto the $X-Y$ (left) and $X-Z$ (right) planes. The galaxy pairs marked with red symbols are
spatially associated in 3D (and thus also in projection). In contrast, the galaxy pairs marked with blue symbols appear to be associated in the $X-Y$
projection, but the $X-Z$ projection clearly demonstrates that they are spatially unassociated. Without accurate redshifts, an observer viewing these
galaxies along the $Z$ axis cannot distinguish between these two types of projected pairs. This figure also illustrates how the spatial information provided
by cosmological simulations is essential for treating the contribution of spatially unassociated projected multiples to the SMG population.

\section{Results} \label{S:results}

\subsection{Cumulative number counts of blended and isolated-disc SMGs}

\begin{figure}
\centering
\includegraphics[width=\columnwidth]{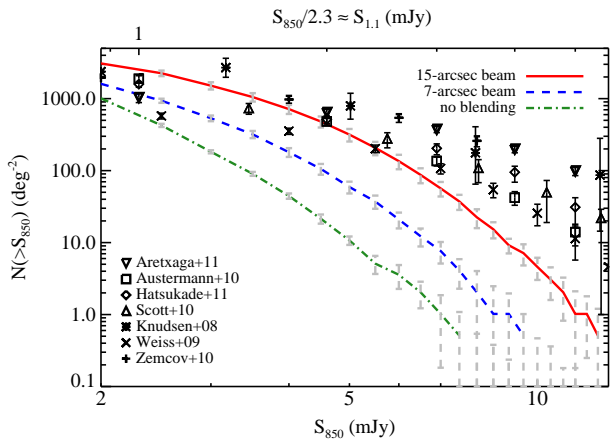}
\caption{Cumulative number counts of model single-dish-detected SMGs (isolated discs and blends of multiple components) versus $S_{850}$ (mJy)
for 15-arcsec (red solid line) and 7-arcsec (blue dashed line) beams.
The number counts predicted when blending is not included are shown for comparison (green dot-dashed line).
All lines denote the median value taken over the eight mock catalogues, and the grey dashed error bars indicate the standard deviation.
The black points with solid error bars are observational data from \citet[][]{Aretxaga:2011},
\citet[][]{Austermann:2010}, \citet[][]{Hatsukade:2011}, \citet[][]{Scott:2010}, \citet*[][]{Knudsen:2008},
\citet[][]{Weiss:2009}, and \citet[][]{Zemcov:2010}; see the legend for the corresponding symbols.
The top axis indicates approximate 1.1-mm flux density values calculated using $S_{1.1} \approx S_{850}/2.3$ (see H13 for details).
The predicted counts agree well with those observed for typical SMGs ($S_{850} \la 6$ mJy), but they are a factor of $\sim 5 - 10$
less (neglecting those of \citeauthor{Aretxaga:2011}; see text for details) for $S_{850} \ga 10$ mJy, which suggests that it is still necessary to
include the SFR elevation caused by starbursts to match the observed counts.}
\label{fig:counts}
\end{figure}

Fig. \ref{fig:counts} shows the SMG cumulative number counts (deg$^{-2}$) versus $S_{850}$
(mJy) predicted for beam sizes of 15 arcsec (which is representative of the resolution of the single-dish telescopes used to determine the
observed counts) and 7 arcsec (for a conservative estimate of the effects of blending) and when blending is not included.
The first two curves include blended SMGs and isolated disc SMGs that are not components of blended sources.
All counts are the median counts taken over the eight mock catalogues, and the grey dashed error bars correspond to the standard deviation.
The black points with solid error bars are observational data (see the caption for details).
The simulation data plotted in this figure are presented in Table \ref{tab:counts}. Observed 1.1-mm counts have
been approximately converted to 850-\micron~counts using $S_{850} \approx 2.3 S_{1.1}$. This conversion is also used to show
approximate $S_{1.1}$ values on the top axis; see H13 for details.

\ctable[
	caption =		{Cumulative number counts for different amounts of blending.\label{tab:counts}},
	center,
	doinside=\small,
	star,
	notespar
]{ccccc}{
	\tnote[a]{850-$\micron$ flux density.}
	\tnote[b]{Approximate 1.1-mm flux density calculated using $S_{1.1} \approx S_{850}/2.3$.}
	\tnote[c]{Cumulative 850-$\micron$ number counts for a beam size of 15 arcsec.}
	\tnote[d]{Same as column (3), but for a 7-arcsec beam.}
	\tnote[e]{Cumulative number counts without the effects of blending included.}
}{
																											\FL
$S_{850}$\tmark[a]	&	$\sim S_{1.1}$\tmark[b]		&	$N_{15''}(>S_{850})$\tmark[c]		&	$N_{7''}(>S_{850})$\tmark[d]	&	$N_{\mathrm{no~blend}}(>S_{850})$\tmark[e] \NN
(mJy)			&	(mJy)					&	(deg$^{-2}$)					&	(deg$^{-2}$)				&	(deg$^{-2}$)	\ML
 2.0     &    0.9    &  $  3084.7  \pm      207.8$    &  $   1606.6 \pm     69.8$   &  $   538.3 \pm     21.1$ \NN
 3.0     &    1.3    &  $  1514.3  \pm      174.8$    &  $    534.2 \pm     48.5$   &  $    98.0 \pm      6.6$ \NN
 4.0     &    1.7    &  $   715.3  \pm      103.0$    &  $    180.1 \pm     25.2$   &  $    24.0 \pm      5.0$ \NN
 5.0     &    2.2    &  $   316.3  \pm       62.5$    &  $     59.2 \pm     11.1$   &  $     5.1 \pm      1.4$ \NN
 6.0     &    2.6    &  $   135.2  \pm       30.7$    &  $     20.9 \pm      5.2$   &  $     1.5 \pm      0.9$ \NN
 7.0     &    3.0    &  $    56.6  \pm       12.9$    &  $      7.7 \pm      2.5$   &  $     0.5 \pm      0.4$ \NN
 8.0     &    3.5    &  $    22.4  \pm        6.6$    &  $      2.0 \pm      2.0$   &  $     < 0.5$ \NN
 9.0     &    3.9    &  $     9.2  \pm        2.1$    &  $      1.0 \pm      0.9$   &  -- \NN
10.0     &    4.3    &  $     4.6  \pm        1.6$    &  $      <0.5$   &  -- \NN
11.0     &    4.8    &  $     2.0  \pm        1.2$    &  --   &  -- \NN
12.0     &    5.2    &  $     1.0  \pm        0.8$    & --   &  -- \NN
13.0     &    5.7    &  $ <0.5 $    &  --   &  -- \LL
}

For the 15-arcsec beam, the predicted counts of blended SMGs are consistent with most of the observed counts for typical SMGs ($S_{850} \la
6$ mJy) but a factor of $\sim 5-10$ less than the observed counts for $S_{850} \ga 10$ mJy.
The \citet{Aretxaga:2011} counts are significantly more discrepant. \citet{Aretxaga:2011}
argue that the origin of the discrepancy between their counts and others observed for similarly sized fields is the impact of galaxy-galaxy weak lensing due to higher than
average matter density along the line of sight. This hypothesis is supported by the association of the excess bright sources with foreground galaxy overdensities
at $z < 1.1$. We do not include the effects of lensing here and thus would not produce this effect if it is the cause.
However, the magnitude of the effect of lensing on the submm counts is still uncertain; see Section \ref{S:limitations} for further discussion.

The difference between the predictions for the 15-arcsec beam (red solid line) and 7-arcsec beam (blue dashed line) indicates that the
total number counts of the blended sources quantitatively depend on the beam size, as one na\"ively expects. This suggests that the
counts observed by telescopes of different resolutions will differ, as is already known \citep[e.g.,][]{Casey:2013}. However, the difference also partially
reflects uncertainty in the number counts predicted by our model because of our simple treatment of blending (see Section \ref{S:limitations}).
Fortunately, there are multiple predictions that are robust to this aspect of the model; we discuss these predictions below.

Note also the significant uncertainty in the $S_{850} \ga 7-10$ mJy counts (depending on the amount of blending),
which originates from sample variance and scatter in the SFR--halo mass relation. This affects all the counts, including the `no blending'
case, because the maximum overdensity sampled varies with the field. It further affects the counts that include blending by altering the probability that two
high-sigma peaks have sufficiently small projected separation to be blended. Thus, even for independent
fields with areas as large as $\sim 2$ deg$^2$, the counts of the brightest sources vary significantly. This effect could be the reason
that the \citet{Karim:2013} and \citet{Smolcic:2012} samples vary in terms of the maximum flux density of non-blended SMGs. For example, for a $\sim 1$ deg$^2$
field, the flux density of the brightest non-blended source observed can vary in the range $S_{850} \sim 6-9$ mJy (note that inclusion of starbursts would
increase these values), and this effect is of course more significant
for smaller surveys. LESS, the survey from which the sources observed by \citeauthor{Karim:2013} were drawn, covers an area of 0.25 deg$^2$,
and the \citeauthor{Smolcic:2012} sources were drawn from the central 0.7 deg$^2$ of the Cosmic Evolution Survey (COSMOS; \citealt{Scoville:2007}) field.
Consequently, because the \citeauthor{Smolcic:2012} survey covered a larger area, the expected flux density for the brightest non-blended SMG in their survey is
greater than for the \citeauthor{Karim:2013} survey.

\subsection{Relative contributions of spatially associated and unassociated components}

\begin{figure}
\centering
\includegraphics[width=\columnwidth]{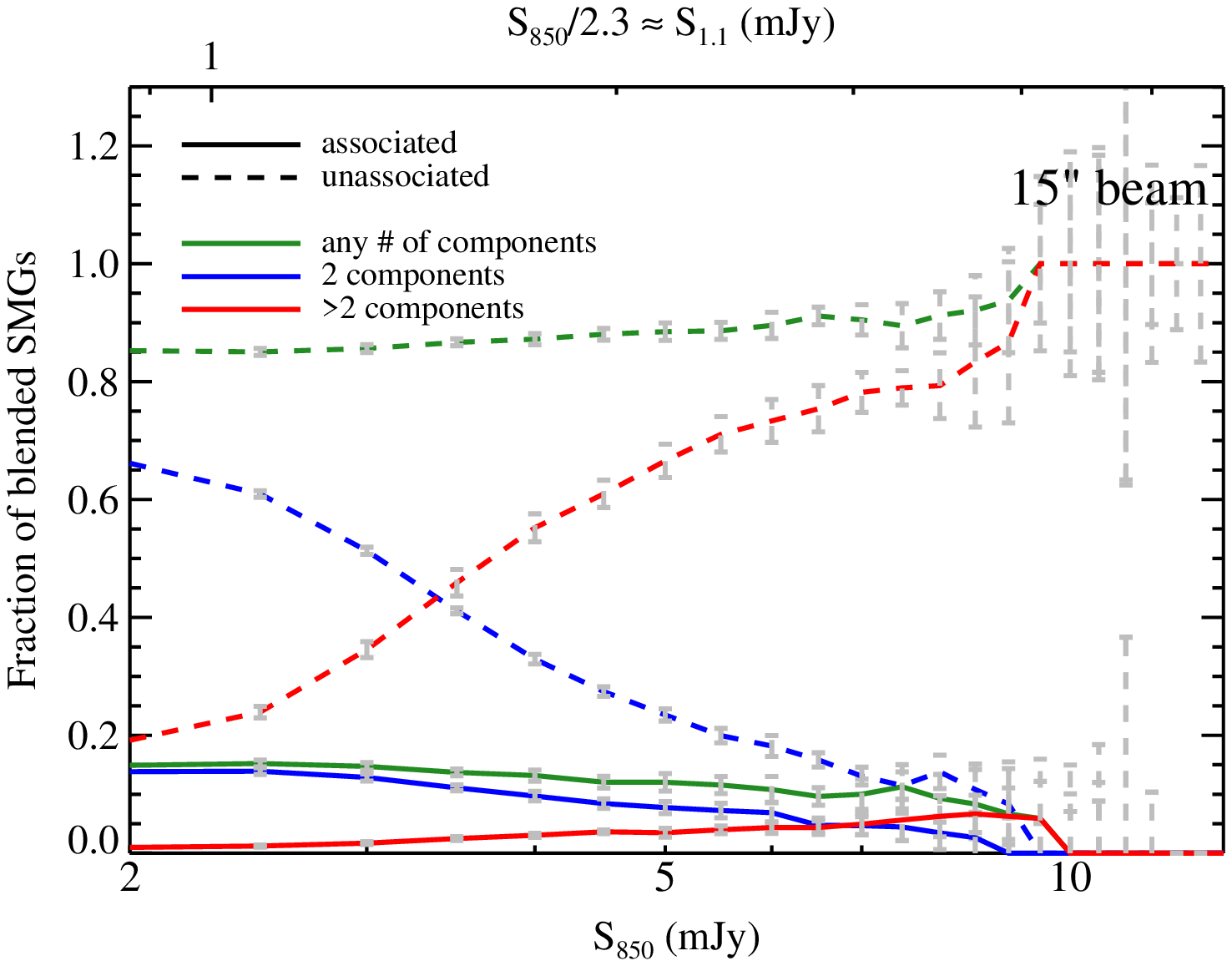} \\
\includegraphics[width=\columnwidth]{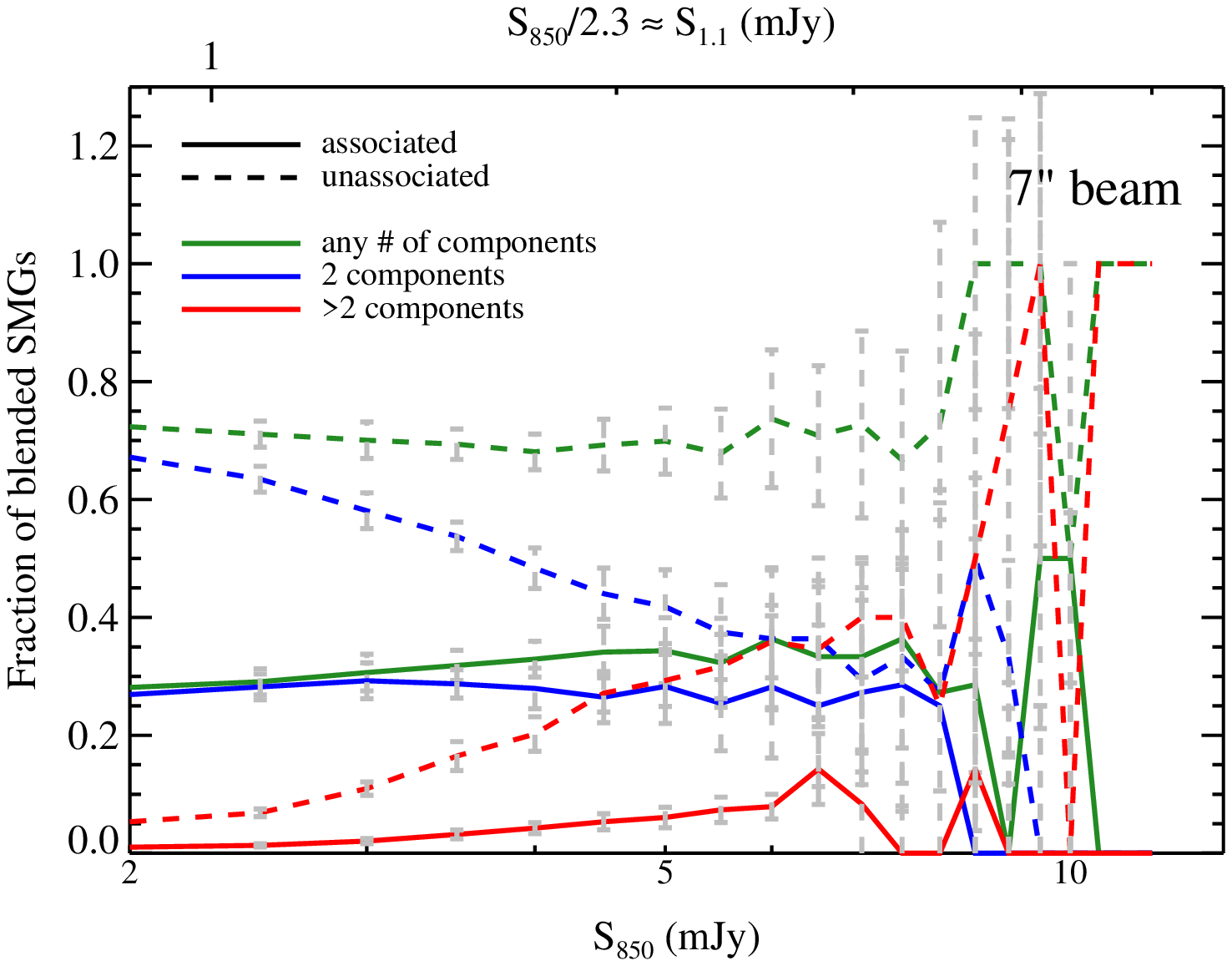} 
\caption{Relative contributions of various subpopulations to the cumulative number counts of blended SMGs for 15-arcsec (top) and 7-arcsec (bottom) beams.
(Non-blended SMGs are ignored here.) The solid lines indicate blended sources for which all components are spatially associated ($\Delta z < 0.02$), and the dashed
lines are for blended SMGs in which at least one component is spatially unassociated ($\Delta z \ge 0.02$). The blue (red) lines correspond to blended SMGs that
are composed of 2 ($>2$) components, and the (dashed) solid green line denotes the contribution of all (un)associated blended SMGs, regardless of the
number of components. The lines indicate the median fraction at a given $S_{850}$ for the eight mock catalogues, and the grey dashed error bars indicate the standard deviation.
At all flux densities plotted, spatially unassociated components contribute to $\ga 50$ per cent of the blended SMGs.
For the 15-arcsec beam, the blended SMGs are dominated by $>2$-component sources in which at least one component is spatially unassociated.
The contribution of spatially unassociated components increases with the beam size.
}
\label{fig:rel_contrib}
\end{figure}

Fig. \ref{fig:rel_contrib} shows the fractional contribution of different subpopulations of blended SMGs to the total blended SMG cumulative number counts
(i.e., the cumulative number counts of that subpopulation divided by the total cumulative number counts of blended sources) versus $S_{850}$. The values are
the medians of the individual ratios for the eight mock catalogues, and the grey dashed error bars indicate the standard deviations.
We consider the spatially associated\footnote{`Spatially associated' is a necessary but not sufficient condition for the individual components
to be interacting galaxies. However, because a classification based only on the difference(s) in the redshifts of the components will be most easily tested with forthcoming
observations, we opt to use this simple criterion rather than a more sophisticated method \citep[e.g.,][]{Moreno:2012}.}
(solid lines) and spatially unassociated (dashed lines) subpopulations, which are defined by $\Delta z < 0.02$ and $\Delta z \ge 0.02$, respectively.
(The justification for this value is presented in Section \ref{S:z_sep}. Here, we simply note that the results are insensitive to the specific cut because the $\Delta z$ distribution
is strongly bimodal.) For both of the aforementioned subpopulations, the fractional contributions of blended
sources composed of two (more than two) components are denoted with blue (red) lines, and the total fractional contributions of associated and unassociated sources
(i.e., independent of the number of components) are shown in green.

One of the robust predictions of our model is that both spatially associated and unassociated blended sources contribute significantly at all flux densities and for the range of beam sizes
explored. Specifically, we expect that assuming a detection limit of $S_{870} > 1$ mJy, $\ga 50$ per cent of blended SMGs contain at least one component
that is spatially unassociated
with the others. The fractional contribution varies with the flux density and beam size and may be sensitive to the treatment of blending, but the significance of the contribution is robust.
For the 15-arcsec beam, the blended SMGs with $S_{850} \ga 3$ mJy are dominated by $>2$-component sources for which at least one component is spatially unassociated with the others.
In contrast, for the 7-arcsec beam, the typical blended SMG in our model is composed of two unassociated galaxies.

Fig. \ref{fig:rel_contrib} also indicates that the relative contribution of spatially unassociated sources is greater for the 15-arcsec beam, especially for the 
$>2$-component sources. This is simply because the greater
beam size increases the likelihood of multiple physically and spatially unassociated galaxies falling with the same beam. In contrast, the relative contribution of spatially associated
blended SMGs is less for the 15-arcsec beam for two reasons: 1. The number of additional physically associated galaxies blended when the beam size is 15 arcsec
is relatively small because the
projected separation of such systems is typically less than the 7-arcsec beam (i.e., $\sim 65$ kpc at $z \sim 2-3$); thus, the 7-arcsec beam is sufficient to blend these sources. 2. Some
sources that are spatially associated for the 7-arcsec beam become classified as spatially unassociated for the 15-arcsec beam because one or more additional, spatially unassociated
galaxies become blended with the spatially associated sources when the beam size is increased to 15 arcsec. For the same reason, the number of sources with $>2$
components is significantly greater when the beam size is 15 arcsec.

\subsection{Redshift separations of the components of blended SMGs} \label{S:z_sep}

\begin{figure}
\centering
\includegraphics[width=\columnwidth]{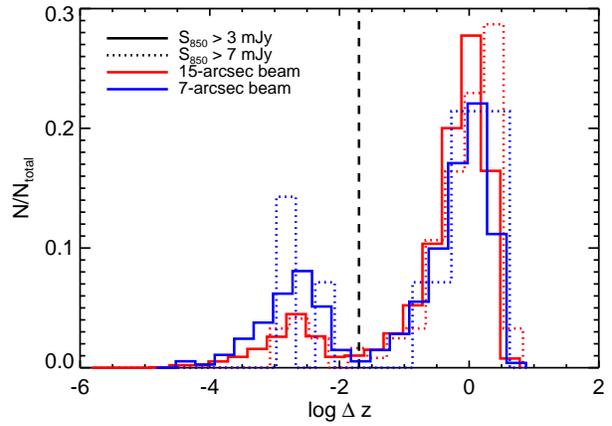}
\caption{The solid (dotted) lines show the distribution of the logarithm of the redshift separation of the individual components (as defined in Eq. \ref{eq:Delta_z}) of the blended submm
sources in one of the mock lightcones with total blended flux density $S_{850} >$ 3 (7) mJy for both the 15-arcsec (red) and 7-arcsec (blue) beams.
The distribution is strongly bimodal, which justifies the $\Delta z$ cut used to separate the spatially associated and unassociated subpopulations; this cut ($\Delta z = 0.02$)
is indicated by the vertical dashed line. The median values are $\Delta z \sim 0.9 - 1.5$, depending on the beam size and flux cut (see Table \ref{tab:med_delta_z}).
Note that the shapes of the distributions are relatively insensitive to the beam size, although for the 7-arcsec beam, a higher fraction of the sources are spatially associated.}
\label{fig:z_distribution}
\end{figure}

\ctable[
	caption =		{Median $\Delta z$ values.\label{tab:med_delta_z}},
	center,
	doinside=\small,
	notespar
]{ccc}{
	\tnote[a]{Median and standard deviation of the $\Delta z$ values for each of the eight mock catalogues of blended SMGs with $S_{850} > 3$ mJy
	and for the beam size specified in column (1).}
	\tnote[b]{Same as column (2), but for $S_{850} > 7$ mJy.}
}{
																	\FL
Beam size	&							&							\NN
(arcsec)		&	$S_{850} > 3$ mJy\tmark[a]	&	$S_{850} > 7$ mJy\tmark[b]	\ML
15			&	$0.99 \pm 0.02$			&	$1.46 \pm 0.16$			\NN
7			&	$0.87 \pm 0.06$			&	$1.00 \pm 0.35$			\LL
}

The final testable prediction that we make concerns the distribution of the redshift separation of the individual components of blended SMGs.
Fig. \ref{fig:z_distribution} shows the distributions of $\log \Delta z$ for blended sources that are brighter than two different flux cuts, $S_{850} > 3$ (solid lines) and 7 (dotted lines) mJy
for both beam sizes (15 arcsec in red and 7 arcsec in blue).
The distributions are strongly bimodal (we interpret this bimodality below), and there is a clear minimum at $\Delta z \sim 0.02$ (marked by the vertical dashed line), which is why we use this value
as the cut to distinguish the spatially associated and unassociated subpopulations. (Using a smaller, more physically meaningful cut, such as $\Delta z = 0.01$, does not
significantly alter the results because of the strong bimodality.) As discussed above, for the 7-arcsec beam, a higher fraction of the sources are spatially associated.

Many of the galaxies in the spatially associated subpopulation ($\Delta z < 0.02$) are also physically associated, i.e., gravitationally bound galaxies that will merge.
The galaxy-pair SMGs (early-stage mergers) discussed in detail in \citet{Hayward:2012smg_bimodality} and H13 are a subset of the spatially associated galaxies.
This category also involves galaxies that are undergoing fly-by encounters but will not merge.
Typically, the spatially associated galaxies are in distinct haloes, and the separations can be as large as $\sim 10$ Mpc (comoving).
The spatially unassociated galaxies ($\Delta z \ge 0.02$) include at least one
component that is simply a chance projection, i.e., is completely unrelated with the other(s). The separations amongst the components of the spatially
unassociated galaxies range from a few Mpc to the box size.

The median and standard deviation of the $\Delta z$ values (calculated from the individual values for each mock catalogue) for the spatially unassociated subpopulations are listed
in Table \ref{tab:med_delta_z}.
Note that the median values are of order 1, which is a natural consequence of the redshift distribution of SMGs in our model:
most SMGs are in the range $z \sim 2-4$, and the median redshift is $\sim 3$.\footnote{At $z \la 2$, the SFRs and (thus submm fluxes) of star-forming galaxies are significantly lower
\citep[e.g.,][]{Noeske:2007a,Daddi:2007,Magnelli:2009,Magnelli:2011,Karim:2011,Whitaker:2012},
a significant fraction of massive galaxies have been quenched \citep[e.g.,][]{Bell:2004,Bell:2007,Faber:2007,Ilbert:2010,Ilbert:2013,Davidzon:2013}, and the volume probed is relatively small;
all of these reasons cause the SMG counts to drop
significantly at $z \la 2$. At $z \ga 4$, the scarcity of massive galaxies and decreased dust content of galaxies causes the counts to decrease. See \citealt{Hayward:2012thesis} and H13 for more details.}
Thus, the typical separation of components is naturally $\sim 1$. The median value of $\Delta z$ is relatively insensitive to the beam size, flux density cut, and mock catalogue used;
thus, it is a robust, testable prediction of our model. The maximum $\Delta z$ varies significantly for the different mock catalogues and ranges from $z \sim 4-6$.

\section{Discussion} \label{S:discussion}

\subsection{Can the brightest SMGs be explained via blending alone?} \label{S:brightest_sources}

The recent observational evidence \citep{Karim:2013,Smolcic:2012} that many of the single-dish SMGs with $S_{870} \ga 12$ mJy are blends
of multiple components was part of the impetus for this work. In particular, we sought to address whether our model could predict a sufficient number of
blended SMGs to account for the observed number counts of sources with $S_{870} \ga 12$ mJy because in the H13 model, such sources are predominantly
starbursts (the models are compared in detail in Section \ref{S:H13_comparison}).

By including the contribution from spatially unassociated components (i.e.,
chance projections), we predict blended SMG number counts (for a 15-arcsec beam) that are consistent with the observed counts within the uncertainties
for typical ($S_{850} \la 6$ mJy) SMGs but a factor of $\sim 5-10$ less than the observed counts for the brightest ($S_{850} \ga 10$ mJy) sources
(ignoring the \citealt{Aretxaga:2011} counts, which are thought to be significantly boosted by weak lensing, as discussed above). The
underprediction at the bright end suggests that inclusion of the SFR elevation caused by starbursts is still necessary to reproduce the brightest sources,
regardless of whether those sources are blends of multiple components.
Furthermore, we must account for the sources with $S_{870} \ga 12$ mJy that are clearly not blended sources, because some examples are known
(see \citealt{Hayward:2013limits} for further discussion); it is very difficult to explain such objects without invoking starbursts induced
by major mergers or disc instabilities. Both issues are deferred to future work, which we outline below.

\subsection{The importance of mock observations for testing models} \label{S:importance_mock_obs}

We have seen that a given single-dish-detected SMG can actually correspond to multiple physical galaxies.\footnote{By `galaxy', we refer to a virialized dark matter halo
that contains baryons; see \citet{Willman:2012} for a more observationally motivated definition.} However, it is often assumed that an SMG corresponds to a single (real or model) galaxy.
Although this assumption is valid for isolated discs and reasonable for late-stage mergers, it is questionable for early-stage mergers, in which the progenitor discs
are well separated and not yet strongly interacting, and definitely invalid for projected multiples.
This error, which is commonly committed by theoreticians but better understood amongst observers, can be partially attributed to the term `submillimetre galaxy',
which is somewhat misleading.\footnote{Perhaps `submillimetre source' is a better term, but the acronym `SMS' is already widely used for another purpose.}

Were this simply a matter of terminology, it would not be a concern, because the astronomical literature is rife with misleading terminology.
However, equating SMGs with individual `galaxies' in simulations or semi-analytical models (SAMs) can lead to \textit{qualitatively} inaccurate conclusions
because of the significant differences between predictions for individual model galaxies (e.g., the `no blending' curve in Fig. \ref{fig:counts}) and predictions
for model SMGs (e.g., the other curves in Fig. \ref{fig:counts}); to make predictions that are classified into the latter category, a model must attempt to
at least crudely mimic observations. This model-comparison problem is not specific to SMGs, of course
\citep[see, e.g.,][]{Lotz:2008,Lotz:2010a,Lotz:2010b,Wuyts:2009b,Wuyts:2010,Scannapieco:2010,Snyder:2011,Snyder:2013},
but it is especially acute for them because 1. observations of SMGs are
more limited than for local and/or less dusty galaxies and 2. the small dynamic range of submm flux density currently probed and the
steepness of the number counts implies that relatively small changes in flux density -- caused by, e.g., blending -- can qualitatively affect the conclusions.
Because of the large beam sizes that are characteristic of far-IR observations, other far-IR-selected populations (e.g., `hot-dust ULIRGs';
\citealt{Chapman:2004,Casey:2009,Casey:2010}) should also be affected by the blending effects discussed here.

\subsection{Implications for the inferred SFRs of SMGs}

Properly treating blending may reconcile the longstanding discrepancy between the far-IR and (sub)mm counts predicted by cosmological simulations and SAMs
and those observed. Furthermore, the possible ubiquity of blended submm sources has an important implication
for interpreting SMG observations: if many of the brightest sources are actually multiple unrelated galaxies, the SFRs inferred under the
assumption that a single submm source corresponds to a single galaxy would overestimate the true SFRs of individual galaxies. Such an overestimate of
the abundance of the most rapidly star-forming galaxies may relieve the tension between the SFRs of the mock SMGs in the cosmological simulation of \citet{Dave:2010}
and those inferred from single-dish observations.

To obtain the true SFRs of individual galaxies, resolved photometry of the individual components is required. In the UV--optical, this process is limited
by the significant attenuation that is characteristic of SMGs. In the IR--(sub)mm, the resolution is the limitation. The ALMA continuum measurements from
the LESS follow-up survey \citep{Karim:2013,Hodge:2013} will help in this regard, but additional high-resolution data in the far-IR, particularly shortward of the
peak of the dust emission, are necessary to accurately constrain the IR luminosity and thus infer the SFRs of the individual components. Therefore, the question
of whether the SFRs of galaxies in cosmological simulations and real galaxies are in tension should be revisited when possible.

\subsection{Comparison with the H13 model} \label{S:H13_comparison}

Although the model presented in this work and that of H13 have some ingredients in common, they differ in the following important aspects: (1) Here, we
directly include information about large-scale structure by using mock galaxy catalogues generated from the \textit{Bolshoi} simulation. In H13, the cosmological
context was obtained solely through the use of merger rates calibrated to cosmological simulations and an assumed stellar mass function. Consequently, no spatial
information was included, and thus chance projections of spatially unassociated galaxies could not be treated. (2) In H13, the effects of blending were
included for early-stage mergers, but in a cruder manner (by assuming a fixed timescale) than in this work. (3) Because the procedure to generate the mock galaxy
catalogues relies on an SFR--halo mass relation (that depends on redshift and includes some scatter), the full SFR enhancement caused by merger-induced starbursts is not
included in this work. In H13, the submm light curves for merger-induced starbursts were directly calculated by performing dust radiative transfer on hydrodynamical
simulations. Consequently, the SFR enhancement in such objects was treated in the H13 model.

The differences described above imply that neither the H13 model nor that presented here can fully treat all proposed SMG subpopulations. Thus, the predictions
of the two models cannot be directly compared. However,
additional insight can be obtained by combining some of their results. Here, we have demonstrated that blending of
unassociated galaxies can significantly boost the submm number counts. Thus, our results reinforce the conclusion of H13 that given a demographically
accurate model galaxy population (i.e., one with stellar mass functions and merger rates consistent with observations), the observed submm counts can be matched
without recourse to variation in the stellar initial mass function. Furthermore, if spatially unassociated multiples were included in the H13 model, it is possible that the model would
overpredict the observed counts. Such an overprediction could imply that the normalisation of the stellar mass function used in the H13 model is too high or that
the H13 simulations are missing some important form of feedback that limits the SFR in starbursts.

\subsection{Limitations of our model} \label{S:limitations}

The strength of the model presented in this work is that it naturally includes information about large-scale structure and thus enables us to physically
characterise the effects of blending for both spatially associated and unassociated galaxies; in contrast, H13 treated only the blending of
the components of early-stage mergers, and the treatment was significantly cruder. However, the method used here naturally also has some limitations.

First, the method by which we assign SFRs to galaxies treats scatter in the SFR--halo mass relation at a given mass and redshift, but it does not
faithfully reproduce the SFR evolution of individual haloes. In particular, the elevation in SFR associated with merger-induced starbursts (i.e., the
high-SFR tail at a given halo mass), which is generically predicted by idealised simulations of major mergers
(e.g., \citealt{Hernquist:1989,Barnes:1991,Barnes:1996,Mihos:1996}; \citealt*{Springel:2005feedback}; \citealt{Cox:2006feedback,Younger:2009}),
is not included. It would be possible to include this effect in an empirically motivated manner \citep[e.g.,][]{Bethermin:2012}. However, because merger-induced
starbursts were treated extensively in H13, and the primary focus of this work is to investigate the contribution of spatially unassociated components
to blended SMGs, we opt to simply neglect merger-induced starbursts here.

Subsets of the spatially associated 2-component and both spatially associated
and unassociated $>2$-component blended SMGs should be undergoing merger-induced starbursts. (The reason that some of the spatially unassociated
$>2$-component blended SMGs should be undergoing starbursts is that in those sources, a subset of the components can be spatially and physically related.)
Because we do not include the elevation in SFR from starbursts, we underestimate the submm fluxes of such objects.
However, because the duty cycle of the starburst phase is significantly less than that of the
`galaxy-pair SMG' phase (H13), the galaxies are not likely to be resolved into separate components during the strong starburst phase,
and not all encounters cause starbursts, the fraction of blended SMGs that are undergoing starbursts should be relatively small.
For more typical spatially associated multiples, which have large separations (i.e., $\ga 10$ kpc),
the SFR elevation should be relatively modest \citep[e.g.,][]{Scudder:2012,Lanz:2013,Patton:2013}.
Consequently, our conclusions regarding the relative contributions of spatially associated and unassociated components to blended SMGs
and the distributions of the redshift separations of the components would not
be significantly affected by the inclusion of starbursts (but the quantitative predictions regarding, e.g., the number counts should be affected, and
the conclusions regarding the brightest sources may differ). Still, there is strong observational evidence that galaxies above the main sequence
(i.e., merger-induced starbursts and gravitationally unstable disc galaxies) contribute to the SMG population \citep[e.g.,][]{Magnelli:2012,Michalowski:2012},
and H13 predicted a significant starburst contribution to the SMG population. Thus, we should and will address this shortcoming of the model presented here in future work.

Furthermore, the treatment of blending is still somewhat crude, albeit significantly less crude than that used in H13 and far superior to simply ignoring blending.
The difference between the predicted counts for the 15- and 7-arcsec beam sizes illustrated by Fig. \ref{fig:counts} indicates that the treatment of blending can
significantly affect some predictions of the model. Although there should certainly be differences in the blended SMG populations identified using telescopes of
different beam sizes, the predictions for a given beam size in the model should not be considered to exactly correspond
to observations from a telescope of the same beam size. Rather, for the typical single-dish SMG surveys, which have been performed with telescopes
with $\sim 15$-arcsec beams, our predictions for the 15- and 7-arcsec beams should be considered somewhat liberal and very conservative
estimates of the number counts of blended SMGs, respectively.

Finally, we have neglected the effects of gravitational lensing on the predicted number counts. As already mentioned above, \citet{Aretxaga:2011}
suggest that galaxy-galaxy lensing is the reason that their counts are significantly higher than those for other fields (see also \citealt{Austermann:2009}).
Furthermore, a population of strongly lensed SMGs has been observed \citep{Vieira:2010}. However, the effect of lensing for sources
with $S_{850} \la 10$ mJy is poorly constrained: estimates of the typical magnification for such sources range from $\sim 10$ per cent
(e.g., \citealt*{Paciga:2009}; \citealt{Lima:2010letter}; \citealt*{Lima:2010}) to of order 10 \citep{Harris:2012}. Thus, we have opted to defer a treatment of lensing to future work.

\subsection{Future work}

In future work, we will combine our fitting function for submm flux density with an updated version of the SAM of \citet{Somerville:2008}
and \citet{Somerville:2012} to produce mock submm maps of the galaxies from the SAM. Then, to make predictions for various telescopes,
we will convolve the maps with the appropriate beams. This approach will enable us to simultaneously treat the isolated disc, merger-induced starburst, and both spatially
associated and unassociated blended SMG subpopulations in order to present a more complete picture of the SMG population.
Furthermore, we will investigate the effects of modifying the prescriptions in the SAM for, e.g., star formation and feedback on the mock SMG population.
The spatial information available in our mock SMG catalogues makes them a natural tool for investigating SMG clustering and the effects of lensing on
submm number counts. Finally, we will complement our semi-analytic approach through the use of state-of-the-art cosmological hydrodynamical simulations
\citep{Vogelsberger:2013,Torrey:2013}.

\section{Conclusions}

We created mock SMG catalogues by generating lightcones from the \textit{Bolshoi} simulation, assigning SFRs and dust masses to the mock galaxies in an
empirically constrained manner, and using a fitting function from previous work to assign submm flux densities. Then, the mock SMG catalogues were used to
investigate the effects of blending of multiple galaxies on the SMG population. Our principal conclusions are the following:
\begin{enumerate}
\item For the 15-arcsec beam, the predicted counts of blended SMGs are consistent with most of the observed counts for typical SMGs ($S_{850} \la
6$ mJy) but a factor of $\sim 5-10$ less than the observed counts for $S_{850} \ga 10$ mJy. The
underprediction at the bright end suggests that inclusion of the SFR elevation caused by starbursts is still necessary to reproduce the brightest sources,
regardless of whether those sources are blends of multiple components.
\item The difference in the maximum flux density of non-blended SMGs in the \citet{Smolcic:2012}, \citet{Karim:2013}, and \citet{Barger:2012} samples is consistent with the
variation in the flux density of the brightest non-blended SMG amongst our mock SMG catalogues. Thus, the difference
may simply be a consequence of sample variance and scatter in the SFR--halo mass relation.
\item Both spatially associated and unassociated blended SMGs contribute significantly to the population:
$\ga 50$ percent of the blended SMGs in our model contain one or more components with $S_{850} > 1$ mJy that are spatially distinct from the other(s).
The fractional contribution depends on both the flux density and beam size.
For a 15-arcsec beam, blends of $>2$ galaxies in which at least one galaxy is spatially unassociated with the others dominate the blended sources with total $S_{850} > 3$ mJy.
\item The distribution of the redshift separations of the components of
blended SMGs is strongly bimodal. For the spatially unassociated subpopulation, the median redshift separation is in the range $\Delta z \sim 0.9-1.5$,
depending on the beam size and flux cut.
\end{enumerate}

We stress that the last two conclusions are bona fide predictions in the sense that (to our knowledge) currently available observational data are insufficient to test them,
but they will be tested as soon as spectroscopic or sufficiently accurate photometric redshifts of the individual components of a sufficient number of blended SMGs are available.

\acknowledgments

We thank Micha{\l} Micha{\l}owski, Ian Smail, and Volker Springel for comments on the manuscript, Jackie Hodge for useful discussion, and the anonymous
reviewer for comments that helped improve the manuscript. CCH is grateful to the organisers of The 30th
Jerusalem Winter School in Theoretical Physics, which stimulated discussions that led to this work, to the Klaus Tschira Foundation for financial support,
and for the hospitality of the Aspen Center for Physics, which is supported by the National Science Foundation Grant No. PHY-1066293.
PSB and RHW received support from HST Theory Grant HST-AR-12159.01-A, provided by NASA through a grant from the Space Telescope Science Institute,
which is operated by the Association of Universities for Research in Astronomy, Incorporated, under NASA contract NAS5-26555.
JRP acknowledges support from a U.S. National Science
Foundation Astronomy and Astrophysics Research Grant (NSF-AST-1010033). JM acknowledges funding from the Canadian Institute for Theoretical Astrophysics and
the Natural Sciences and Engineering Research Council of Canada (Discovery Grant; PI: S.~L.~Ellison).
\\

\footnotesize{
\bibliography{std_citations,smg}
}

\label{lastpage}

\end{document}